# AlN interlayer to improve the epitaxial growth of SmN on GaN (0001)

S. Vézian,[1,a] B. Damilano,[1] F. Natali,[2] M. Al Khalfioui[1,3] and J. Massies[1]

[1] Centre de Recherche sur l'Hétéro-Épitaxie et ses Applications (CRHEA), Centre National de la Recherche Scientifique, Rue Bernard Gregory, 06560 Valbonne, France.

[2] MacDiarmid Institute for Advanced Materials and Nanotechnology, School of Chemical and Physical Sciences, Victoria University of Wellington, PO Box 600, Wellington, New Zealand.

[3] Université de Nice Sophia Antipolis, Parc Valrose, 06102 Nice Cedex 2, France.

**Abstract:**

An *in situ* study of the epitaxial growth of SmN thin films on Ga-polar GaN (0001) templates by molecular beam epitaxy is reported. Using X-ray photoelectron spectroscopy we found that Ga segregates at the surface during the first stages of growth. We showed that the problem related to Ga surface segregation can be simply suppressed by growing a few monolayers of AlN before starting the SmN growth. This results in a significant improvement of the crystallinity of SmN thin films assessed by X-ray diffraction.

---

[a] Corresponding author : sv@crhea.cnrs.fr

1. Introduction

The recent demonstration that a new class of ferromagnetic materials - the rare-earth nitrides (REN) - are epitaxy-compatible with group III-nitrides (GaN, AlN and InN) [1,2,3,4] - the technologically important nonmagnetic semiconductor family for the fabrication of white and blue light emitting diodes and transistors - has raised interest not only for semiconductor-based spintronics but also for the possibility of enhancing the efficiency of GaN-based light emitting diodes [5,6,7]. Success in obtaining REN thin films epitaxially grown on wurtzite (0001) oriented AlN or GaN surfaces has been central in getting a better understanding of their fundamental properties, in particular demonstrating, for most of them, their intrinsic ferromagnetic semiconducting nature with a wide variety and complementary magnetic properties across the series [5]. So far, GdN and SmN thin films, typically of the order of tens nanometers thick, have been the most studied compounds of the series, with several reports on the effect of the growth parameters (growth temperature, RE-nitrogen flux ratio…) on the structural and electronic properties [1,2,8,9].

Developing heterojunction device structures based on these two nitride families will rely on the understanding and the ability to control, at the atomic scale, the interface structure and chemical stability. Hitherto these aspects have not been studied in depth, and in this paper we propose to investigate the very first stages of the epitaxial growth of SmN on Ga-polar GaN(0001), the orientation used for technological applications. We show that gallium segregates at the surface during growth, presumably resulting from an exchange reaction between Sm and Ga at the SmN/GaN interface. A method to overcome the problem of Ga surface segregation is presented based on the insertion of an AlN interlayer before the growth of SmN.

2. Experimental

Samples were grown in a Riber molecular beam epitaxy (MBE) system equipped with a RHEED gun (STAIB instruments 25 kV). Al, Ga and Sm are provided by conventional solid sources. Molecules of ammonia (NH$_3$) and pure nitrogen (N$_2$) were used as nitrogen precursor for the growth of GaN/AlN and SmN, respectively. It is worth mentioning that the growth of SmN, like GdN and some other REN, can be carried out under pure N$_2$ atmosphere thanks to the catalytic dissociation of molecular nitrogen by RE atoms on the growing surface [2]. The samples studied here were grown on 1.5 μm thick n-type (Si doping at 3×10$^{18}$ cm$^{-3}$) Ga-polar GaN(0001) grown on a Si(111) substrate [10]. The SmN layers were grown at a substrate temperature of 400 °C under N-rich conditions, with a beam equivalent pressure (BEP) of 2.7×10$^{-5}$ Torr and 5×10$^{-8}$ Torr for N$_2$ and Sm respectively, corresponding to a growth rate of 0.1 μm/h. *In situ* scanning tunneling microscopy (STM) and X-ray photoelectron spectroscopy (XPS) measurements were performed at room temperature. XPS was carried out using a Mg Kα (hv=1253.6 eV) non-monochromated X-ray source, equipped with a 7 channels hemispherical analyzer, using a pass energy of 10 eV. All measurements were taken at the normal incidence of the sample. For the X-ray diffraction (XRD) measurements, 100 nm thick SmN layers were capped with GaN (thickness of 100–150 nm) in order to prevent decomposition in air [2].

3. Results and discussion

    3.1.1 SmN growth on bare GaN surface

Prior to the SmN growth, the *in situ* reflection high-energy electron diffraction (RHEED) pattern shows the typical 2×2 surface reconstruction of Ga-polar GaN (0001) at low temperature (below 550°C) [11]. When the growth of SmN starts on GaN, the diffraction pattern changes drastically from sharp streaks to a weak and diffuse background. After a few monolayers (about 4-5 MLs, 1 ML = 0.29 nm), some circles appear which are characteristic of the diffraction by polycrystalline film (Fig. 1(a)). Then the pattern slowly evolves to a spotty diagram after the deposition of 10-15 nm of SmN (see Fig. 1(b)). Note that the spotty pattern observed is similar to the one reported for GdN, with double spots, linked to twinned domains in the face-centred cubic (fcc) structure of REN, along the GaN [1-210] azimuth (Fig. 1(c)). Fig. 2(a) shows an *in situ* STM image of ~13 nm thick SmN layers grown on GaN where the surface morphology is similar to the one reported for GdN layers grown on AlN surfaces [8]. The root mean square (rms) roughness is about 1 nm.

We now turn to XPS chemical surface analysis. We proceed by several cycles of SmN deposition followed by *in situ* XPS measurements (Fig. 3). The SmN thickness ranges from 0 to ~13nm. The Sm 4d, Ga 3p and N 1s core level spectra are recorded as a function of the SmN coverage. Concerning the Sm 4d peak (see Fig. 3(a)), its intensity increases when the SmN growth proceeds, as expected. This peak is structured by three contributions at 135.0 eV, 131.4 eV and 128.8 eV, respectively, typical of the RE series due to the $4d^9 4f^n$ interaction [12]. We have also measured the energy position of the N 1s core level peak as a function of the SmN coverage. A shift in energy from 397.0 eV to 396.1 eV is observed when increasing the SmN coverage (Fig. 3(b)). This shift has a chemical origin and can be explained as follows. Before growing SmN, the N 1s component corresponds to the GaN layer, where each nitrogen atom is bonded to four gallium atoms. Once the growth of SmN proceeds, a second component is superimposed due to the change in the crystal structure, from wurtzite to rocksalt, where each nitrogen atom is now surrounded by six samarium atoms. Therefore, the observed shift is related to the change in the binding energy of the N 1s core level.

The behaviour of the Ga 3p spectrum as a function of the SmN coverage displayed in Fig. 3(c) is more tricky to understand: the Ga 3p peak intensity decreases but does not disappear. The remaining presence of the Ga 3p peak, even for thick SmN layers, suggests either SmN islanding growth or Ga segregation at the SmN surface during growth. Interestingly, the Ga 3p peak (a doublet with 3p 1/2 and 3p 3/2 components) is shifted during the growth of SmN from 107.7 eV and 104.3 eV (GaN starting surface) to 106.9 eV and 103.5 eV (after the growth of ~13 nm of SmN), while no significant energy shift is observed for the Sm 4d. Such a shift of about 0.8 eV towards low binding energy when increasing the SmN coverage suggests that Ga is no more bound to nitrogen as in GaN. Thus we believe that the remaining presence of the Ga peak for thick SmN layers is not related to parts of the GaN surface not entirely covered by the SmN layer due to islanding growth, but rather indicates a Ga surface segregation phenomenon. This is well confirmed by considering the Ga LMM Auger transition spectrum evolution with SmN growth reported in Fig. 4(a). When growing SmN directly on GaN, a second set of components is rapidly superimposed to the initial one coming from GaN, shifted from 4 eV to the high kinetic energy side, and becomes predominant after 3.9 nm growth. Such Auger line chemical shift of 4 eV has been reported for metallic gallium on GaN [13,14], confirming the Ga segregation already mentioned above.

To go further, we have calculated the integrated areas under the core level peaks reported in Fig. 3. Fig. 5(a) shows the integrated intensity of Sm 4d peaks as a function of the SmN thickness (red circles). In addition, the integrated intensity of Ga 3p peaks (red circles) is reported as a function of the SmN thickness in Fig. 5(b). Experimental curves can be fitted using the Beer-Lambert relationship:

$$I = I_0 exp(-d/\lambda) \qquad (1)$$

where $I_0$ is the photoelectrons intensity emitted for all depth greater than $d$ and $\lambda$ is the attenuation length (AL) [15]. Considering a two dimensional (2D) surface layer of thickness $d$ (SmN in our case), the intensity of electron emitted from the substrate (across the surface layer) is given by Eq. (1) where $I_0$ is the intensity without the surface layer. On the other hand, to obtain the expression for the signal coming from a thin SmN surface layer of thickness $d$, the Beer-Lambert equation must be integrated (between 0 and $d$) and becomes:

$$I = I^\infty \left[1 - exp(-d/\lambda)\right] \qquad (2)$$

where $I^\infty$ is the intensity from the "bulk" material (i.e. with $d$ infinite).

Applying these equations to our experimental data, we found AL values $\lambda$ of (11 ± 4) nm for Ga 3p and (3.0 ± 0.3) nm for Sm 4d. In the case of Sm 4d, we took 12.6 nm for the value of $I^\infty$ (corresponding to 4 $\lambda$, a reasonable approximation). If we neglect the elastic collisions in first approximation, the attenuation length (AL) and the inelastic mean free path (IMFP) can be used interchangeably. By using the method of Tanuma et al. [16] to calculate the IMFP for Sm 4d peak (kinetic energy of 1122 eV) we found 3.2 nm which is in good agreement with the AL extracted from Fig. 5(a) and validates our approximation of 2D surface layer, i.e. no islanding growth. On the other hand, for the Ga 3p peak (kinetic energy of 1180 eV), IMFP of 3.3 nm is obtained, i.e. significantly lower than the AL (11 nm) determined experimentally. This is fully consistent with the presence of Ga at the surface of the sample. In turn, this would mean that GaN bond-breaking occurs in the very first stages of the SmN growth. It is worth mentioning that, even at room temperature, a chemical reaction takes place at the interface when Sm (Ref. 17) or Ni (Ref. 18) are deposited onto GaN. In both cases, it is concluded that Ga is released at the interface. In the present work, the growth of SmN being performed at 400°C, the reactivity at the surface should be enhanced. Owing to the strong reactivity of Sm towards nitrogen, an exchange reaction between Sm and Ga at the GaN surface can result. As Ga, contrary to Sm, does not react with molecular nitrogen (even at temperature far above 400°C), we believe that "free" Ga is formed at the interface and is segregated during the subsequent SmN growth.

### 3.1.2 SmN growth using AlN interlayer

At this stage, we can wonder if such interfacial reaction occurs if the SmN growth is performed on AlN instead of GaN surface. Keeping in mind that the final objective is to grow SmN epitaxial layer with the best possible structural quality on GaN, a limited thickness of AlN should be deposited. Indeed, AlN should be elastically strained on GaN in order to avoid the formation of dislocations. The 2.4% lattice mismatch between AlN and GaN results in a critical thickness for plastic relaxation as thin as 12 ML (~3 nm) [19]. Interestingly, the first stages of SmN growth on an AlN interlayer on GaN show a behavior significantly different to the one corresponding to the SmN growth on bare GaN, and this from only 2 MLs of AlN (1 ML = 0.25 nm). At the onset of the growth, the RHEED pattern does not disappear as in the case of growth on the bare GaN: the starting AlN diagram (streaky) coexists with a spotty one, located at lower spacing distance which corresponds to a larger lattice parameter as expected for a SmN (111) surface ($a = 3.56$ Å for SmN, compared to 3.19 Å for GaN and pseudomorphic AlN). This indicates a sharp interface formation. In addition, there is no circle formation characteristic of polycrystalline phase. After the growth of typically 3-4 MLs, only the spotty pattern related to the SmN layer remains. This pattern is almost identical to the one obtained after the deposition of 15 nm of SmN grown on bare GaN reported in Fig. 1(b).

A STM image of a 13 nm thick SmN layer grown on AlN surface (Fig. 2(b)) shows that the surface roughness is similar to the one obtained for SmN grown on GaN (rms of ~1 nm). However, there is a surface morphology difference, with lower size island dispersion and more pronounced facets when SmN is grown on AlN surface.

Integrated area under core level peaks versus the SmN thickness for Sm 4d and Al 2p peaks deduced from XPS measurements on SmN grown on 8 ML AlN interlayer on GaN are reported in Fig. 5 (blue circles). The Sm 4d data agree well with the one obtained for SmN growth on bare GaN, indicating that the growth rate of SmN is the same for the two different experiments (Fig. 5(a)). Considering now the Al 2p peak (kinetic energy of 1150 eV) coming from the underlying AlN, no shift is observed when the SmN growth proceeds (Fig. 3(d)), contrary to the behavior observed for the Ga 3p peak during the SmN growth on GaN (Fig. 3(c)). In addition, an experimental AL of $(3.6 \pm 0.5)$ nm is deduced from its intensity attenuation by the SmN layer (Fig. 5(b)), in agreement with calculated IMFP of 3.24 nm. Also, the Ga signal from the underlying GaN layer rapidly decays. As shown in Fig. 4(b), the Ga LMM Auger transitions are no more observed after the growth of ~10 nm of SmN, as expected. Furthermore, there is no shift of the Ga Auger lines during the growth of SmN. These results evidence that the Ga segregation is suppressed when an AlN interlayer is used.

Therefore, when growing SmN on AlN surface there is no evidence of Al surface segregation or interfacial reaction between Sm and AlN surface. This can explain the difference of the RHEED pattern evolution during the very first stages of the growth of SmN on GaN and AlN surfaces: the odd behavior observed on bare GaN surface is related to the reaction of Sm on GaN leading to a diluted and complex interface.

We also observed the presence of two components in the N 1s peak (not shown), similarly to what we reported above for the growth on bare GaN surface. This peak shifts to lower binding energy during SmN deposition from 397.7 eV to 396.1 eV. Thus chemical shifts of 0.9 eV and 1.6 eV are observed during the growth of SmN on GaN and AlN, respectively. Even though the energy shift of the core levels measured by XPS also integrate final state effects phenomena like intra-atomic trapping or surface relaxation, it has been shown that initial state effects (chemical bonding) are often the dominant factors [20]. Indeed, the N 1s energy shift of 0.7 eV observed between GaN and AlN well agrees with the difference in the cohesive energy per bond of GaN and AlN which is 2.20 eV and 2.88 eV, respectively [21]. An estimate of the cohesive energy per bond in SmN of $(1.3 \pm 0.1)$ eV is therefore found by subtracting the N 1s chemical shift to the cohesive energy for both type of samples. This result is consistent with first principles calculation performed on some REN, and in particular for GdN: a cohesive energy per bond of about 1 eV is deduced from the cohesive energy per atom value (5.9 eV) [22]. We note that, to the best of our knowledge, no cohesive energy value for SmN has been reported.

From RHEED, XPS as well as STM experiments reported above, we expect some difference in the structural properties of SmN depending on the starting growth surface. Therefore, we have performed X-ray diffraction on 100 nm thick SmN layers directly grown on GaN and on an AlN interlayer inserted before the SmN growth. The thickness of the AlN interlayer has been varied from 2 ML to 20 ML. The rocking curve ($\omega$-scan) of the SmN (111) diffraction peak has been recorded and its full width at half-maximum (FWHM) is reported in Fig. 6. The FWHM decreases of about an order of magnitude when the AlN interlayer increases up to 8 ML and then slowly increases. The latter is probably related to the plastic relaxation of AlN on GaN when the AlN thickness exceeds ~12 ML, resulting in the formation of dislocations impacting the crystalline quality of the SmN overlayer. The key result is however the strong FWHM decrease observed even for an AlN interlayer of only 2 ML. As AlN has the same structure as GaN and is elastically strained, i.e. has the same in-plane parameter, the only reason for the improvement of the SmN epitaxial layer structural quality is due to the suppression of the interface reaction at the SmN/GaN interface and the resulting Ga surface segregation.

4. Conclusion

In summary, we have studied the first stages of the epitaxial growth of SmN on GaN (0001) by RHEED, XPS and STM. RHEED indicates the formation of a diluted and complex SmN/GaN interface. In turn, XPS spectra recorded as a function of the SmN growth show that Ga segregates at the surface of the growing layer. The insertion of only a few monolayers of AlN at the SmN/GaN interface is sufficient to observe a sharp interface formation by RHEED. This confirms that a specific reaction occurs between Sm atoms and the GaN surface: Sm-N bonds are formed to the detriment of Ga-N ones, resulting in the release of Ga which segregates at the surface of the SmN growing layer. Finally, the FWHM of X-ray diffraction rocking curve is significantly improved by the presence of an AlN interlayer at the SmN/GaN interface, clearly demonstrating the key role of the interface chemistry on the structural quality of SmN epitaxial layers on Ga-polar GaN (0001).

Acknowledgement

The authors thank Prof. Joe Trodahl for fruitful discussions and valuable suggestions. We acknowledge support from GANEX (ANR-11-LABX-0014). GANEX belongs to the public funded "Investissements d'Avenir" program managed by the French ANR agency. We acknowledge funding from the Marsden Fund (Grant No. 08-VUW-1309), and the MacDiarmid Institute for Advanced Materials and Nanotechnology, funded by the New Zealand Centres of Research Excellence Fund.

**Figure captions**

Fig.1 : RHEED patterns recorded along the [1–210] azimuth of GaN(0001) after the growth of (a) 2 nm and (b) 15 nm of SmN on GaN. (c) Geometrical arrangement of diffraction spots along the ⟨1–10⟩ direction for a fcc single crystal (blue circles). A second fcc crystal rotated by 180° (red circles) corresponding to twins domain is superimposed.

Fig.2 : 100×100 nm$^2$ STM images of 13 nm thick SmN grown at 400°C (a) directly on GaN and (b) on a 8 ML thick AlN interlayer grown on GaN. Acquisition parameters are 0.35 nA and +2 V (empty states).

Fig.3 : XPS spectra taken at the (a) Sm 4d, (b) N 1s, (c) Ga 3p and (d) Al 2p core level regions. (a), (b) and (c) from sample directly grown on GaN while (d) from sample grown on 8 ML thick AlN interlayer grown on GaN.

Fig. 4 : XPS spectra taken at the Ga LMM Auger electron peaks region (a) for a sample directly grown on GaN and (b) for a sample grown on 8 ML thick AlN interlayer grown on GaN. Red and blue lines are guides for the eyes and indicate peak positions corresponding to metallic Ga and GaN respectively.

Fig.5 : Integrated area under core level peaks versus SmN thickness for (a) Sm 4d and (b) Ga 3p and Al 2p peak components. Red and blue circles correspond to SmN grown on bare GaN and on 8 ML thick AlN interlayer grown on GaN, respectively. Black solid lines are the fits using Eq. (2) for (a) and Eq. (1) for (b).

Fig.6 : Full width at half-maximum (FWHM) of the out of plane (111) rocking curve of 100 nm thick SmN layer as a function of the AlN interlayer thickness. AlN thickness is expressed in monolayer (ML), 1 ML = 0.25 nm. The black line is only a guide for the eye.

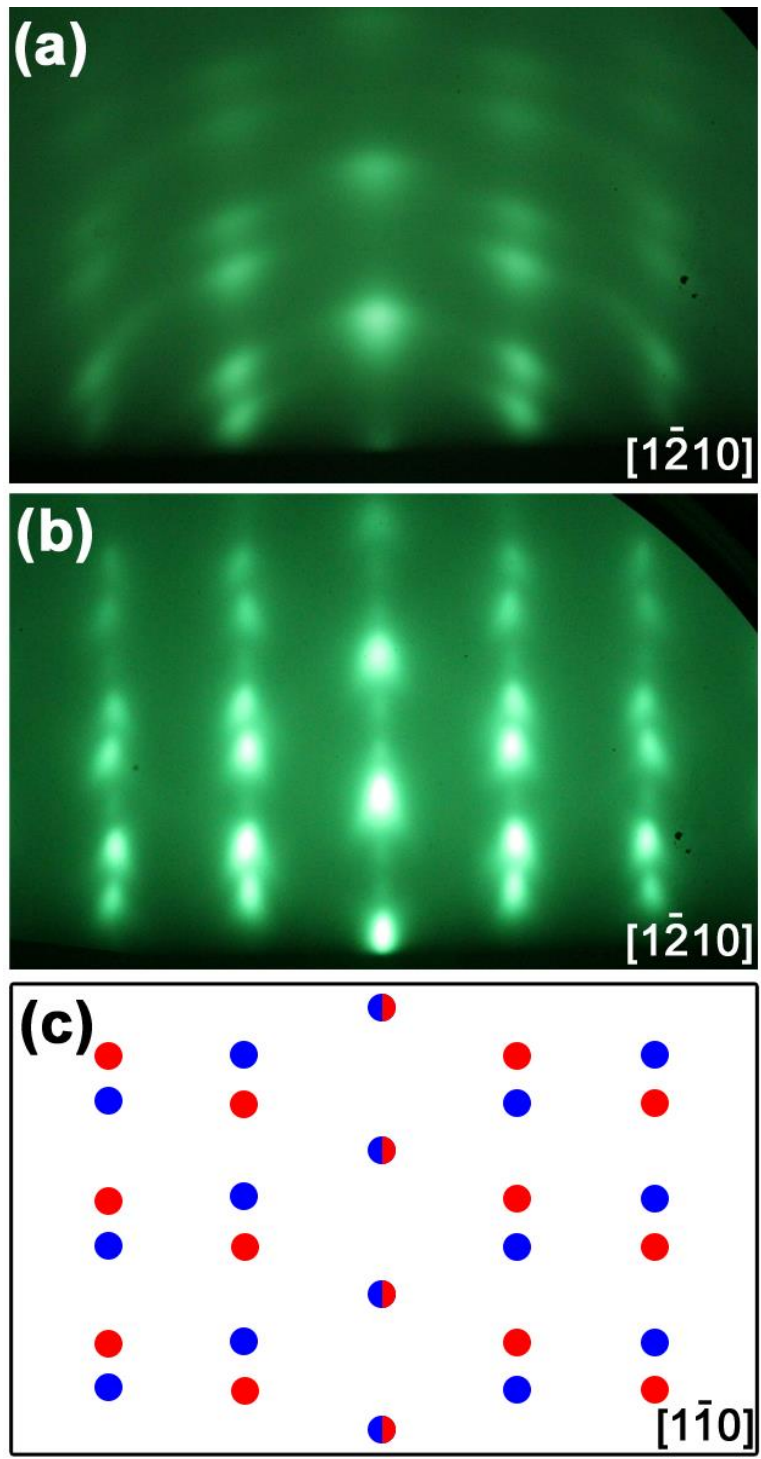

*Figure 1.*

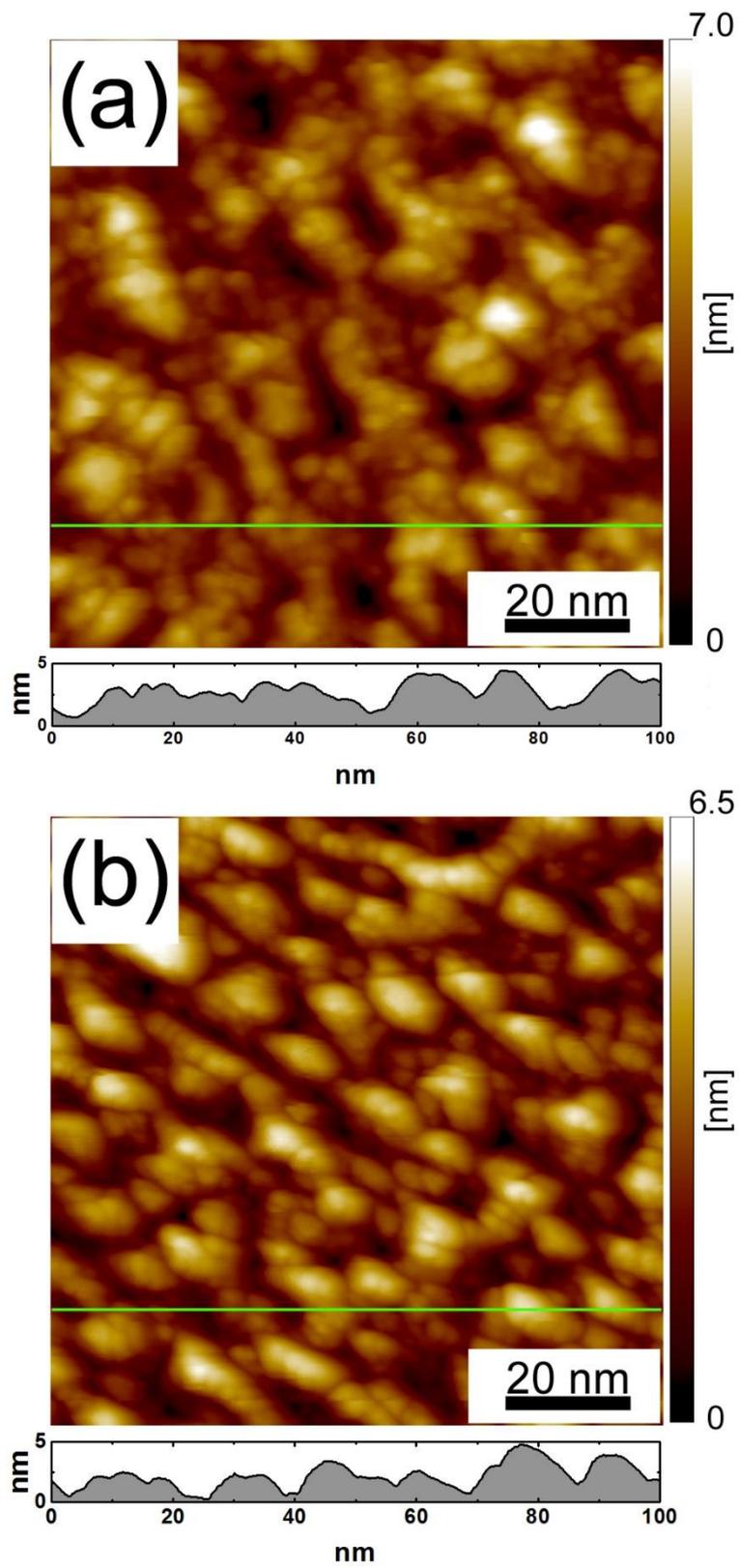

*Figure 2.*

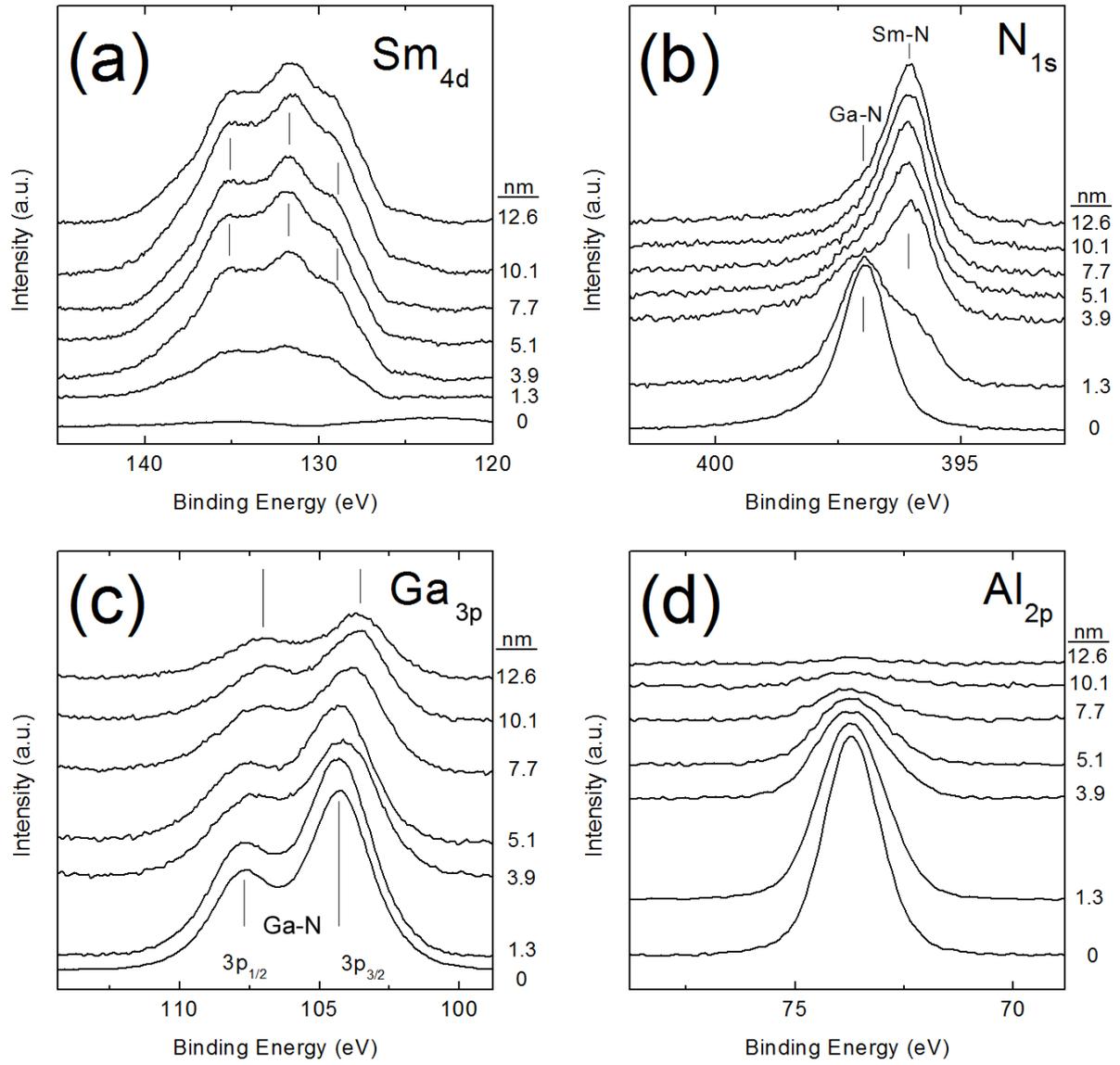

*Figure 3.*

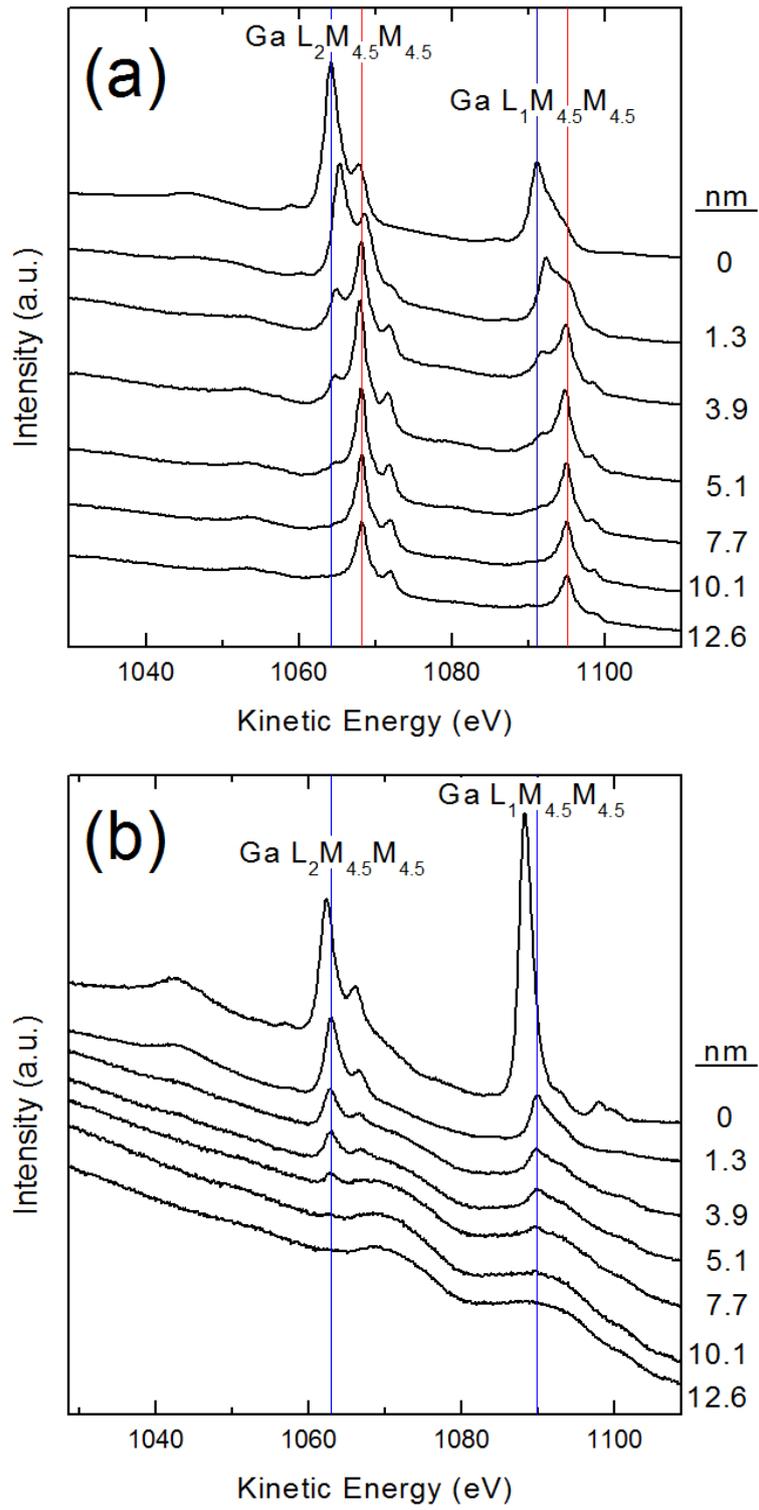

*Figure 4.*

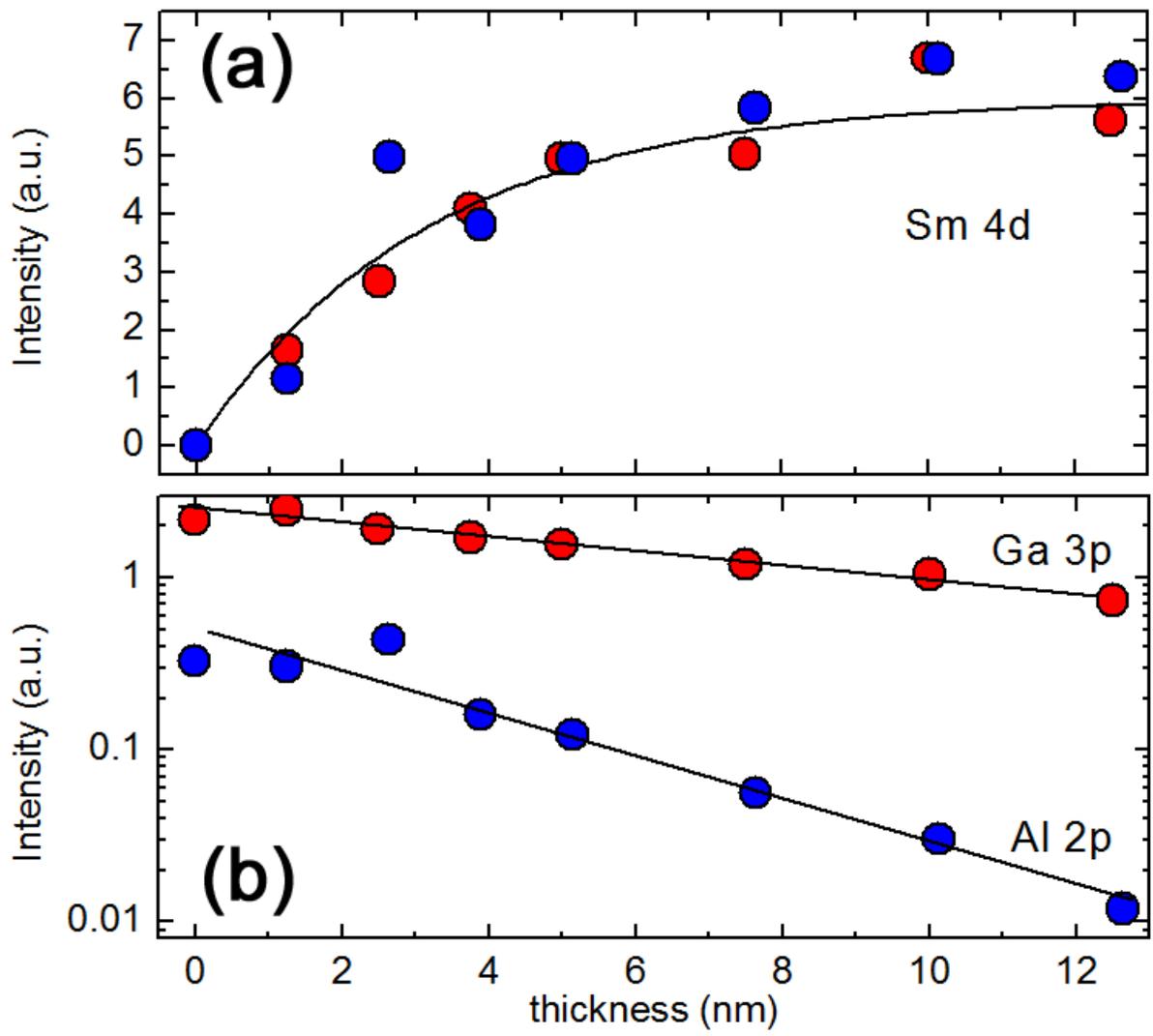

*Figure 5.*

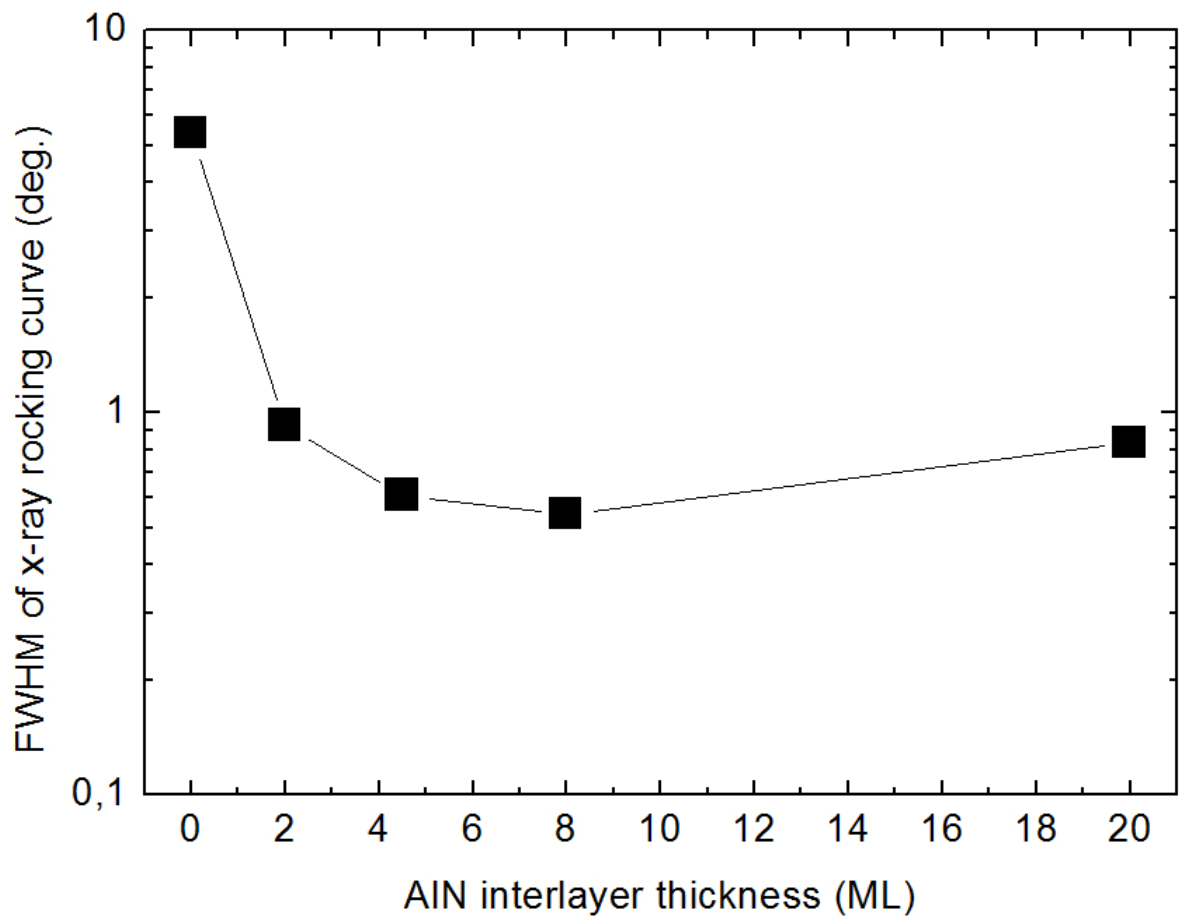

*Figure 6.*